\documentclass[11pt,a4paper]{article}
\usepackage{jheppub}
\usepackage[]{tensor}
\def\qqd{\ , \quad}
\def\dd{\partial}
\newcommand{\ctg}{\textrm{ctg}\,}
\def\sL{\sqrt{\lambda}}
\def\0{\nonumber}
\def\be{\begin{eqnarray}}
\def\ee{\end{eqnarray}}

\DeclareSymbolFont{lettersA}{U}{txmia}{m}{it}
 \DeclareMathSymbol{\rr}{\mathord}{lettersA}{"92}
 \DeclareMathSymbol{\cc}{\mathord}{lettersA}{"83}

\title{A new look at hidden conformal symmetries\\ of black holes}

\author[a]{Edgardo Franzin}
\author[a,b]{and Ivica Smoli\'c}

\affiliation[a]{International School for Advanced Studies (SISSA/ISAS),\\
Via Bonomea 265, 34136 Trieste, Italy}
\affiliation[b]{Theoretical Physics Department, Faculty of Science, University of Zagreb,\\
p.p.~331, HR-10002 Zagreb, Croatia}

\emailAdd{edgardo.franzin@gmail.com}
\emailAdd{ismolic@phy.hr}

\abstract{We propose a different approach to the analysis of symmetries in the near-horizon region of black holes. The idea is presented here for spherically symmetric black holes, for which we have shown that the generators of hidden symmetries can be extracted by solving the conformal Killing equation under certain assumptions. Our result is in agreement with that obtained in literature by the analysis of the wave equation in the near-horizon region.}


\arxivnumber{1107.2756}

\begin{document}

\begin{flushright}
SISSA/35/2011/EP, \, ZTF-11-07
\end{flushright}

\maketitle

%%%%%%%%%%%%%%%%%%%%%%%%%
\section{Introduction}%%%
%%%%%%%%%%%%%%%%%%%%%%%%%
\label{s:intro}

The black hole entropy arises in a semiclassical formulation of general relativity~\cite{BCH:1973,Beken1973} and it is given by the well-known Bekenstein-Hawking formula,
\begin{equation*}
S = \frac{A}{4}
\end{equation*}
where $S$ denotes the black hole entropy and $A$ its event horizon area.

The first microscopic derivation of this semiclassical result has been provided in string theory by counting microstates~\cite{SV:1996} for black holes characterized by a near-horizon region with an AdS$_3$ factor.
Afterwards Strominger showed~\cite{Strominger:1997} that any consistent and unitary quantum theory of gravity, containing those particular black holes as solutions, must reproduce the entropy essentially in the same way,
with no need for specific details of string theory.
Brown and Henneaux~\cite{Brown:1986} had already noticed that a consistent completion of quantum gravity on AdS$_3$ has to be described by a two-dimensional conformal field theory (CFT). In addition Carlip~\cite{Carlip} and Solodukhin~\cite{Solodukhin} 
have shown that, given plausible boundary conditions on the horizon, the algebra of deformations at the black hole horizon is the Virasoro algebra.

A considerable amount of effort has been put forth to
unveil the hidden conformal symmetries in the near-horizon region of black holes.
Recently, a correspondence has been conjectured between CFT and Kerr black holes (see~\cite{BKLS:2011} for a review).
In the original formulation~\cite{GHSS:2009} of the Kerr/CFT correspondence the duality is between extreme Kerr black holes and a chiral two-dimensional CFT\@.
In this case, quantum gravity has been studied in the near-horizon
region of the extreme Kerr black hole, which is analogous to the
AdS$_3$ region considered in previous works. Given consistent boundary
conditions, it has been shown that the algebra of the asymptotic
symmetry generators is again the Virasoro algebra.
In the case of non-extreme Kerr black holes a different approach~\cite{CMS:2010} is proposed.
The idea is to consider the wave equation for a scalar massless field in the Kerr background and then studying it in a particular limit. It is possible to define a set of vector fields such that they obey the $SL(2,\rr)$ algebra and their Casimir reproduces the scalar wave equation operator. Recent results have shown that this approach can also be applied to Kerr-Newman~\cite{Wang:2010} and Schwarzschild~\cite{BCK:2011} black holes. Convincing arguments for universality of hidden conformal symmetries for a very large class of higher-dimensional black holes have been given in~\cite{Krishnan:2010}.

\medskip
In this paper we propose a different approach based on directly solving the conformal Killing equation, without introducing auxiliary scalar or other fields. The idea is to restrict the analysis to some specific submanifold of the spacetime, by assuming particular form of the (conformal) Killing vectors. Furthermore, we may allow \emph{some} components of the metric $g_{ab}$ to be \emph{a priori} undefined functions $g_{ab}(x)$. Then, by solving the conformal Killing equation, we search for conditions upon
\begin{enumerate}
\item[a)] the form of the functions $g_{ab}(x)$ and
\item[b)] the region of spacetime,
\end{enumerate}
which allow for the appearance of some additional Killing vectors, not present in the underlying spacetime. In the general case it is expected that the metric components $g_{ab}(x)$, obtained as solutions to the field equations, do not satisfy these conditions, so a Taylor series expansion, in an appropriate region of the spacetime, may be required.

\medskip
The remainder of the paper is organized as follows. In section~\ref{s:confKill} we analyze the solutions to the conformal Killing equation, related to the $t$-$r$ submanifold of the spherically symmetric spacetime, and we discuss the conditions under which certain additional Killing vectors may appear. In section~\ref{s:nhsymm} we make particular choices of integration constants for the Killing vectors, in order to define generators which reproduce a $SL(2,\rr)$ algebra, related to the hidden symmetry. Eventually, in section~\ref{s:finrem} we make some final remarks about the advantages of our approach and its relation to some other approaches in literature. In Appendix~\ref{s:app} we present components of Christoffel symbol used in the calculation.

\vspace{0.5cm}

%%%%%%%%%%%%%%%%%%%%%%%%%%%%%%%%%%%%%%%%%%%%%%%%%%%%%%%
\section{Analysis of the conformal Killing equation}%%%
%%%%%%%%%%%%%%%%%%%%%%%%%%%%%%%%%%%%%%%%%%%%%%%%%%%%%%%
\label{s:confKill}

In order to simplify the analysis we limit ourselves to the case of spherically symmetric black hole solutions in $D$ spacetime dimensions. Next, we assume that the metric has the form
\be\label{met}
ds^2 = -f(r)\,dt^2 + \frac{dr^2}{f(r)} + r^2\,d\Omega_{D-2}^2
\ee
written in Schwarzschild coordinates (see Appendix~\ref{s:app}). Birkhoff's theorem is known to be valid for a spherically symmetric solution in a wide range of theories~\cite{Zegers,BCPPS:2011b}, so we assume that the function $f$ is strictly $r$-dependent. We also assume that $f$ is not a constant function. Moreover, it seems reasonable to assume that $g_{tt} \, g_{rr} = -1$~\cite{Jacobson:2007}, at least to make the analysis as simple as possible at this point. For the Schwarzschild 4-dimensional black hole with mass $M$, the function $f(r)$ has an explicit form and we shall use the following notation,
\be
f_{\textrm{Sch}}(r) = 1 - \frac{2M}{r}
\ee
All the comments about the 4-dimensional Schwarzschild solution made throughout this paper are straightforward to generalize to higher-dimensional cases or to the charged, Reissner-Nordstr\"om case.

\medskip
Leaving the angular part (and its $SO(D\!-\!1)$ symmetry) aside, we focus on the isometries pertaining to the $t$-$r$ submanifold. To this end we limit our investigation to Killing vectors~$\xi^\mu$ with vanishing angular components and independent of angular coordinates,
\be
\xi^t = \xi^t(t,r) \qqd \xi^r = \xi^r(t,r) \qqd \xi^i = 0 \quad (2 \le i \le D-1)
\ee
In order to make our procedure more ``flexible'' we start from the conformal Killing equation (see~\cite{Wald}, Appendix~C),
\be
\pounds_\xi \, g_{ab} = \sigma(x) \, g_{ab}
\ee
where $\sigma(x)$ is some function. More explicitly, we have the following equation
\be
\nabla_{\!(\mu} \xi_{\nu)} = \sigma(x) \, g_{\mu\nu} \label{cKE}
\ee
Equation (\ref{cKE}) evaluated for the metric (\ref{met}) splits up into a system of differential equations. Most of those including angular components $\xi_i$ are trivially satisfied (under proposed assumptions), except for the $ii$-components, which read
\be
r\,f\,\Pi(i)\,\xi_r = \sigma\,r^2\,\Pi(i) \label{cKEi}
\ee
so that
\be
\sigma(t,r) = \frac{f(r)}{r} \, \xi_r(t,r) \label{mu}
\ee
The remaining equations are listed below\footnote{A dot indicates derivative with respect to $t$, whereas a prime indicates derivative with respect to $r$.}
\be
&&\dot{\xi}_t - \frac{ff'}{2} \, \xi_r = -\frac{f^2}{r} \, \xi_r \label{cKE1}\\
&&\xi'_r + \frac{f'}{2f} \, \xi_r = \frac{1}{r} \, \xi_r \label{cKE2}\\
&&\dot{\xi}_r + \xi'_t - \frac{f'}{f} \, \xi_t = 0 \label{cKE3}
\ee
At this point we make some approximations, by going to the near-horizon region of the black hole where $f(r) \approx 0$. This allows us to neglect RHS terms\footnote{RHS terms are, by construction, proportional to the function $\sigma(x)$, and we expect them to be subleading in the near-horizon limit, corresponding to $\sigma(x) \to 0$ or ``conformal $\to$ ordinary'' Killing equation limit.} in equations (\ref{cKE1}) and~(\ref{cKE2}). The consistency of this approximation will be justified \emph{a posteriori}, once we know the explicit solutions to the simplified equations,
\be
&&\dot{\xi}_t - \frac{ff'}{2} \, \xi_r = 0 \label{KE1}\\
&&\xi'_r + \frac{f'}{2f} \, \xi_r = 0 \label{KE2}\\
&&\dot{\xi}_r + \xi'_t - \frac{f'}{f} \, \xi_t = 0 \label{KE3}
\ee
We can first integrate (\ref{KE2}) since it is a purely $r$-dependent equation,\footnote{If we had left the RHS term in (\ref{cKE2}), the solution (\ref{solxir}) would have modified by an $r$-factor. It is easy to check that for such solution the RHS term in (\ref{cKE2}) is again subleading in the near-horizon approximation.}
\be
\xi_r(t,r) = \frac{A(t)}{\sqrt{f(r)}} \label{solxir}
\ee
where we have introduced a $t$-dependent function $A(t)$. Putting this result back into (\ref{KE1}) and (\ref{KE3}), we have
\be
&& \dot{\xi}_t - \frac{f'\sqrt{f}}{2} \, A = 0 \label{KE1b}\\
&& \frac{\dot{A}}{\sqrt{f}} + \xi'_t - \frac{f'}{f} \, \xi_t = 0 \label{KE3b}
\ee
Time derivative of (\ref{KE3b}) together with (\ref{KE1b}), lead to
\be\label{Af}
\ddot{A} + \frac{1}{4} \, \Big( 2ff'' - (f')^2 \Big) A = 0
\ee
It is possible to make separation of variables, with constant $\lambda$,
\be
\frac{\ddot{A}}{A} = \lambda = -\frac{1}{4} \, \Big( 2ff'' - (f')^2 \Big)
\ee
which leads to the following system of differential equations,
\be
&&\ddot{A} - \lambda A = 0 \label{Adeq}\\
&&2ff'' - (f')^2 + 4\lambda = 0 \label{fdeq}
\ee
We first assume that $\lambda \ne 0$ and afterwards make a comment about the $\lambda = 0$ case. The solution to (\ref{Adeq}) is well known,
\be\label{ASol}
A(t) = \alpha e^{\sL t} + \beta e^{-\sL t}
\ee
where $\alpha$ and $\beta$ are integration constants.
The differential equation (\ref{fdeq}) is nonlinear, but can be reduced to the following form,
\be
(f')^2 - 4\lambda = Bf \qqd B = \textrm{const.}
\ee
There are two possible cases,
\be
f(r) &=& \pm \,2\sL \, r + C \qqd B = 0, \label{fSol1}
\ee
and
\be
f(r) &=& \frac{B}{4} \, r^2 + \frac{BC}{2} \, r + \frac{(BC)^2 - 16\lambda}{4B} \qqd B \ne 0, \label{fSol2}
\ee
where $C$ is a new integration constant. Hence, solutions to equation (\ref{fdeq}) are functions $f(r)$ which are either quadratic or linear polynomial (constant functions have been excluded from the beginning).

\medskip
Going back to the $\xi_t$ component, we can first integrate the differential equation (\ref{KE1b})
\be
\xi_t(t,r) = \frac{\sqrt{f}f'}{2} \int^t A(t') \, dt' + g(r)
\ee   
where it has been assumed that any integration constant coming from $t'$-integral, multiplied by $\sqrt{f}f'/2$, is already absorbed into $g(r)$. This means that we can use (\ref{Adeq}) to write this result in somewhat more elegant form,
\be\label{xit}
\xi_t(t,r) = \frac{\sqrt{f}f'}{2\lambda} \, \dot{A}(t) + g(r)
\ee
If we insert (\ref{xit}) into equation (\ref{KE3b}), we get
\be
g'(r) - \frac{f'}{f} \, g(r) - \frac{\dot{A}}{4\lambda\sqrt{f}} \, \Big( (f')^2 - 2ff'' - 4\lambda \Big) = 0
\ee
The third term vanishes due to (\ref{fdeq}), so a direct integration leads to
\be
g(r) = K f(r) \qqd K = \textrm{const.}
\ee

Finally, we can write the components of the most general Killing vector obtained using near-horizon approximations,
\be
\xi^r &=& g^{rr} \xi_r = \sqrt{f(r)} \Big( \alpha e^{\sL t} + \beta e^{-\sL t} \Big)\\
\xi^t &=& g^{tt} \xi_t = -K - \frac{1}{2\sL} \, \frac{f'(r)}{\sqrt{f(r)}} \left( \alpha e^{\sL t} - \beta e^{-\sL t} \right)
\ee
Three arbitrary constants, $\alpha$, $\beta$ and $K$, allow us to write three independent Killing vectors. Going back to the point where we made the initial approximation for the near-horizon region, i.e.~in equations (\ref{cKE1}) and (\ref{cKE2}), it is easy to check that the RHS terms have been discarded in a consistent manner. Also, we note that in the near-horizon limit the function $\sigma(t,r)$ goes to zero, so that the conformal Killing equation (\ref{cKE}) reduces to the ``ordinary'' Killing equation.

\medskip
The function $f(r)$ obtained from the field equations is rarely a polynomial (most notable exceptions being Minkowski and (A)dS spacetimes). In the non-polynomial case the only solution to (\ref{Af}) is the ``trivial'' one with $A(t) = 0$, which implies that we have only one Killing vector satisfying conditions from above, given by
\begin{equation*}
\xi^a = -K \left( \frac{\dd}{\dd t} \right)^{\!a}
\end{equation*}
However, if $f(r)$ is an analytic function in some neighbourhood of the horizon, we can make a Taylor expansion around that point up to quadratic terms in $r$ and then identify additional near-horizon symmetries. Since $f(r_\text{h}) = 0$, zeroth order expansion does not make much sense in the near-horizon region; some additional remarks about first-order (linear) expansion will be given below. But, one might ask what happens if we proceed with the series and include higher order terms. A possible answer is that by doing so, we add more and more details from the underlying spacetime (also, one has to move further away from the horizon in order to make higher order terms relevant) and in this way we destroy hidden symmetries in the reduced near-horizon region. This observation also provides us with some further insight into the limitations of this kind of identification of hidden (conformal) symmetries.

Let us now consider the 4-dimensional Schwarzschild black hole as an example. Function $f_{\textrm{Sch}}(r)$ is analytic in the neighbourhood of the horizon at $r = r_\text{h} = 2M$, so we can make the following Taylor expansion,
\be\label{SchTaylor}
f_{\textrm{Sch}}(r) &=& \frac{1}{2M} \, (r - 2M) - \frac{1}{4M^2} \, (r - 2M)^2 + \mathcal{O}((r-2M)^3) = \0\\
&=& -\frac{r^2}{4M^2} + \frac{3r}{2M} - 2 + \mathcal{O}((r-2M)^3)
\ee
from which we can immediately read off the parameters in (\ref{fSol2}),
\be
B = -\frac{1}{M^2} \qqd C = -3M \qqd \lambda = \frac{1}{16M^2} \0
\ee

\medskip
Before proceeding with the identification of the hidden symmetry we add a comment about the $\lambda = 0$ case. In this instance the solution to (\ref{Adeq}) is a linear function $A(t)$ whereas solutions to (\ref{fdeq}) are again quadratic or linear polynomials $f(r)$. However, it turns out that in this case it is not possible to consistently define generators out of Killing vectors, so that they close into an algebra. For this reason we discard this case as irrelevant for our discussion.  

\vspace{0.5cm}

%%%%%%%%%%%%%%%%%%%%%%%%%%%%%%%%%%%%
\section{Near-horizon symmetries}%%%
%%%%%%%%%%%%%%%%%%%%%%%%%%%%%%%%%%%%
\label{s:nhsymm}

We start by defining three independent Killing vectors, $\widetilde{H}_{+1}$, $\widetilde{H}_0$ and $\widetilde{H}_{-1}$, with the following choice of constants,
\be\label{Htilde}
\widetilde{H}_{+1} = \xi|_{\alpha = i, \, \beta = 0, \, K=0} &=& i e^{\sL t} \left( \sqrt{f} \, \dd_r - \frac{1}{2\sL} \, \frac{f'}{\sqrt{f}} \, \dd_t \right)\\
\widetilde{H}_0 = \xi|_{\alpha = 0 = \beta, \, K \ne 0} &=& - K \dd_t\\
\widetilde{H}_{-1} = \xi|_{\alpha = 0, \, \beta = -i, \, K=0} &=& -i e^{-\sL t} \left( \sqrt{f} \, \dd_r + \frac{1}{2\sL} \, \frac{f'}{\sqrt{f}} \, \dd_t \right)
\ee
Their commutators are given by
\be\label{tildcomm}
[\widetilde{H}_0, \widetilde{H}_{\pm 1}] = \mp K \sL \, \widetilde{H}_{\pm 1} \qqd [\widetilde{H}_{+1}, \widetilde{H}_{-1}] = -\frac{f''}{K\sL} \, \widetilde{H}_0
\ee
If one uses a linear approximation for the function $f(r)$ this algebra simplifies since $\widetilde{H}_{+1}$ and $\widetilde{H}_{-1}$ commute, and it is isomorphic to the algebra of the 2-dimensional Poincar\'e group $ISO(1,1)$. For quadratic approximation $f''(r) = B/2$ is a constant and the commutation relations (\ref{tildcomm}) are related to the commutation relations of the $SL(2,\rr)$ algebra,
\be\label{SL2R}
[H_0, H_{\pm 1}] = \mp i H_{\pm 1} \qqd [H_{+1}, H_{-1}] = 2i H_0
\ee
In order to properly normalize generators we make the choice,
\be
K = \frac{i}{\sL} \quad ,
\ee
followed by redefinition
\be
H_0 = \widetilde{H}_0 \qqd H_{\pm 1} = \gamma \, \widetilde{H}_{\pm 1}
\ee
with $\gamma = \sqrt{B}/2$. Finally, introducing the surface gravity $\kappa = 2f'(r_\text{h})$, it is easy to show that $\lambda = \kappa^2$, and we can write these three generators in a somewhat simplified form,
\be\label{Hgen}
H_{+1} &=& \frac{i}{\gamma} \, e^{\kappa t} \left( \sqrt{f} \, \dd_r - \frac{1}{2\kappa} \, \frac{f'}{\sqrt{f}} \, \dd_t \right)\\
H_0 &=& - \frac{i}{\kappa} \, \dd_t\\
H_{-1} &=& -\frac{i}{\gamma} \, e^{-\kappa t} \left( \sqrt{f} \, \dd_r + \frac{1}{2\kappa} \, \frac{f'}{\sqrt{f}} \, \dd_t \right)
\ee
It is straightforward to verify that these generators obey the $SL(2,\rr)$ algebra (\ref{SL2R}). Moreover, the corresponding Casimir operator for this algebra is given by
\be
\mathcal{H}^2 &=& -H_0^2 + \frac{1}{2} \, (H_{+1} H_{-1} + H_{-1} H_{+1}) = \0\\
&=& \dd_r \left( \frac{1}{\gamma^2} \, f \, \dd_r \right) + \frac{1}{\kappa^2} \left( 1 - \frac{1}{4\gamma^2}\,\frac{(f')^2}{f} \right) \dd^2_t 
\ee 

\medskip
Finally, we look at the 4-dimensional Schwarzschild solution, where (using quadratic approximation) we have
\be
\sL = \kappa = \frac{1}{4M} \qqd B = -\frac{1}{M^2} \qqd \gamma = \frac{i}{2M} \0
\ee
In order to relate our generators to those from a recent paper~\cite{BCK:2011}, we have to make a formal limit to the ``asymptotic infinity'' (defined by $r \to \infty$) of the near-horizon region,
\be
\sqrt{f} \to \gamma \sqrt{\Delta}
\ee
which suffice to complete the identification.

\vspace{0.5cm}

%%%%%%%%%%%%%%%%%%%%%%%%%%%%%%%%%%%%%%%%%%%%%
\section{Final remarks and open questions}%%%
%%%%%%%%%%%%%%%%%%%%%%%%%%%%%%%%%%%%%%%%%%%%%
\label{s:finrem}

In this paper we have shown how to identify the hidden isometries of the near-horizon region of a spherically symmetric static black hole, using the reduced form of the conformal Killing equation. In the case of Schwarzschild black hole these additional Killing vectors obey the $SL(2,\rr)$ commutation relations, in accordance with the recent result~\cite{BCK:2011}, obtained by the wave equation approach. The same conclusion can now be extended to spherically symmetric static black holes, where the function $f(r)$ in metric (\ref{met}) is analytic in some neighbourhood of the event horizon.

\medskip
We find two beneficial features of the procedure presented in this paper. First, there is no need to guess the form of generators or some peculiar choice of coordinates which would ease the identification of hidden symmetries. Killing vectors are obtained as solutions to the conformal Killing equation constrained by near-horizon approximations. Second, our method presents a clear insight into the limitations of such procedure: the identification of these additional symmetries is valid as long as we remain in the near-horizon region, so that terms of the order $\mathcal{O}((r - 2M)^3)$ in the Taylor series for the metric function $f(r)$ remain subleading.

\medskip
It will be interesting to see if it is possible to use this approach to reproduce hidden conformal symmetry for the Kerr black hole, obtained in~\cite{CMS:2010} by the wave equation approach. Another open problem is whether there is some connection between the delicate choice of the second-order expansion in Taylor series and the choice of boundary conditions used in~\cite{Carlip,Solodukhin,GHSS:2009}.

\vspace{0.5cm}

%%%%%%%%%%%%%%%%%%%
\acknowledgments %%
%%%%%%%%%%%%%%%%%%%

We would like to thank prof.~Loriano Bonora for encouragement and inspiring discussions. Also, we would like to thank SISSA for hospitality and financial support. E.F.~wrote this paper during his postgraduate fellowship at SISSA\@.
I.S.~would like to acknowledge the financial support of CEI Fellowship Programme CERES and support by the Croatian Ministry of Science, Education and Sport under the contract no.~119-0982930-1016.

\vspace{1.0cm}

\appendix

%%%%%%%%%%%%%%%%%%%%%%%%%%%%%%%%%%%%%%%%%%%%%%
\section{Metric and Christoffel components}%%%
%%%%%%%%%%%%%%%%%%%%%%%%%%%%%%%%%%%%%%%%%%%%%%
\label{s:app}

Metric (\ref{met}) is written in Schwarzschild coordinates
\be
x^0 = t \qqd x^1 = r \qqd x^i = \theta^i \quad (2 \le i \le D-1)
\ee
where $t$ is time, $r$ is radial and $\theta^i$ are angular coordinates, such that
\be
0 \le \theta^i < \pi \quad \textrm{for} \quad i = 2, \dots, D-2 \quad \textrm{and} \quad 0 \le \theta^{D-1} < 2\pi \0
\ee
Note that the last coordinate $\theta^{D-1}$ is more frequently denoted by $\phi$. Throughout the paper indices $i$, $j$ and $k$ are used for angular coordinates ($2 \le i,j,k \le D-1$).

\medskip\noindent
Components of the metric (\ref{met}) can be written in a compact way, using auxiliary function
\be
\Pi(i) = \left\{ \begin{array}{ccl}  1 & , & i = 2 \\ & & \\ \prod_{k=2}^{i-1} \, \sin^{2} \theta^k  & , & i \ge 3 \end{array} \right.
\ee
\noindent
so that
\be
g_{00} = - f(r) \qqd g_{11} = \frac{1}{f(r)} \qqd g_{ii} = r^{2} \, \Pi(i)
\ee

\medskip\noindent
Using this, it is straightforward to calculate the components of Christoffel symbol,
\begin{gather}
\Gamma^{0}_{01} = \frac{f'}{2f} \qqd \Gamma^{1}_{00} = \frac{f f'}{2} \qqd \Gamma^{1}_{11} = - \frac{f'}{2f} \qqd \Gamma^{1}_{ii} = -r f \, \Pi(i) \ , \0\\
\Gamma^i_{1i} = \frac{1}{r} \qqd \Gamma^i_{ij} = \Gamma^i_{ji} = \ctg{\theta^j} \quad \textrm{for} \quad i \ge j+1 \label{Cristoff}\\
\Gamma^i_{jj} = -\ctg \theta^i \prod_{k=i}^{j-1} \sin^{2}{\theta^k} \quad \textrm{for} \quad j \ge i+1 \0
\end{gather}
where a prime indicates derivative with respect to $r$.

\vspace{0.5cm}

%%%%%%%%%%%%%%%%%%%%%%%%%%%%%%%%%%%%%
%   t h e b i b l i o g r a p h y   %
%%%%%%%%%%%%%%%%%%%%%%%%%%%%%%%%%%%%%
\bibliographystyle{JHEP}
\bibliography{references}
%\nocite{*}

\end{document}